\documentclass[prb,twocolumn,showpacs,preprintnumbers,amsmath,amssymb]{revtex4-1}

\usepackage{graphicx,rotating}
\usepackage{dcolumn}
\usepackage{bm}
\usepackage{epsfig}
\usepackage{longtable}
\usepackage{color}

\begin{document}

\title{New memory devices based on the proton transfer process.}  

\author{Ma\l{}gorzata~Wierzbowska\email{wierzbowska@ifpan.edu.pl}}

\affiliation{Institut of Physics, Polish Academy of Science (PAS),
Al. Lotnik\'ow 32/46, 02-668 Warszawa, Poland}

\date{\today}

\begin{abstract}
Memory devices operating due to the fast proton transfer (PT) process are  
proposed by means of the first-principles calculations.
Writing an information is performed using the electrostatic potential 
of the scanning tunneling microscopy (STM). 
Reading an information is based on the effect of the local magnetization 
induced at the zigzag graphene nanoribbon (Z-GNR) edge - saturated with 
oxygen or the hydroxy group - and can be realized with the use of 
the giant magnetoresistance (GMR), 
magnetic tunnel junction (MTJ) or spin-transfer torque (STT) devices.
The energetic barriers for the hop-forward 
and -backward processes can be tuned by the distance and potential of the STM tip.
Thus, enabling to tailor the non-volatile logic states. 
The proposed system enables very dense packing of the logic cells and
could be used in the random access and flash memory devices.
\end{abstract}


\maketitle

\section{Introduction}

Memory devices require densely packed systems, operating with high speed and
involving stable processes for non-volatility. 
They also should not dissipate much heat \-- for
the energy saving reasons. 
Various mechanisms are used for switching the logic states in the memory devices.
A review on the magnetic memory devices can be found in the Noble lecture
by Fert.\cite{noble} Merged magnetic and superconducting switching devices have been proposed
recently.\cite{josephson} Organic resistive memory devices with a high on/off ratio are also
available,\cite{org} and these systems can be 3D integrated.\cite{3D-org,3D-org2}  
Other 3D integrated systems can be constructed for the optical memory devices based 
on the photochromic molecules.\cite{3D-opt} 
The basic idea of the devices proposed in this work is to use the electric field
for writing the logic state and the local magnetization for reading information.  
 
It is already known, that in the zigzag graphene nanoribbon with the edges saturated 
with different species, the local magnetic moment arises.\cite{mag-1,mag-2}
The above phenomenon, was already 
planned to be used in the spintronic devices for the spin-current
filtering.\cite{spin-filter,filter2}
Depending on a choice of the species used for the saturation
and their alternation along the edge, the local moments show up at the both edges 
or only at one of them. The whole edge might be magnetic or only some segments   
show nonvanishing local moments. The effect is very close to the edge or extends
over a few bond lengths.
If the local magnetization is well pronounced, even for frequent alternation of
different species along the edge \-- or when the segments are short, 
i.e. the logic cells are small 
\-- then the highly packed memory devices can be built. 
Using the above phenomenon in a connection with the 
giant magnetoresistance-based 
detector\cite{gmr} enables very fast reading of information.

The information encrypted at the Z-GNR edge depends on its
saturation. Therefore, one can saturate edges with the oxygen atoms and the OH groups 
\-- since they adsorbe via the double- and single-bonds, respectively. 
Thus, writing information could be attained with the use of 
the proton transfer (PT) phenomenon. The hydrogen atom can move from 
the edge saturated with OH to the STM tip,
which is decorated with some small aromatic molecule saturated with O; 
and vice versa (from the STM tip to the Z-GNR).   
Binding/releasing the hydrogen atom to/from the molecule attached to 
the STM appex is similar to the vertical and horizontal transport of small
molecules in the surface-engeneering process.\cite{penseta-1,penseta-2}
With the difference that the molecular transport occurs
directly from the metallic tip, while in our case, the molecule is always
attached to the tip and only H jumps.   

The PT process between the aromatic molecules (or the graphene edge) 
is fast \-- a few picoseconds or even below.\cite{fast,fast1,fast2}
The total techological process, however, would be restricted by the speed
of the alternation of the electric field \-- which is of order of gigahertz 
in the scanning electron microscopy (SEM). On the other hand, this part
could be replaced by new terahertz technology. 
The energy barriers for the jump forward and backward are tunable
by the distance and potential of the STM tip.
Hence, the high speed and non-volatile memory cells could be designed.
The processes involved in the proposed system did not require the electric
current flow. Therefore, the device would not be overheated and the energy dissipation
would be low, as well as the scaling limits (characteristic for the resistive memories)
would not be a problem.\cite{scal1,scal2}

The above writing mechanism, 
namely the proton transfer within the O-H...O and O-H...N groups, 
is a very popular process in the organic systems. 
This is the basic mechanism responsible for the helical structure of DNA double strand 
and the secondary structure of proteins. 
The molecules subjecting to the PT phenomenon have been proposed as building blocks of 
the electronic devices, such as: rectifiers and transistors,\cite{elec1}
photo- and field-switches,\cite{elec2,elec3} high frequency devices,\cite{NDR}
and the optical data storage.\cite{mem1,mem2} 

The binding or releasing the hydrogen atom, in the organic systems, 
induces a change of the single- and double-bond pattern, which can be mapped in the STM images.  
The PT process is also correlated with a change of the dipole moment \--
which, however, cannot be directly measured.
Recent progress in material science is fast. Thus, the molecular prototypes based 
on the PT process might be utilized in the memory devices similar to the Z-GNR edge,
in the nearest future. 

\begin{figure*}[ht]
\centerline{
\includegraphics[scale=0.43,angle=0.0]{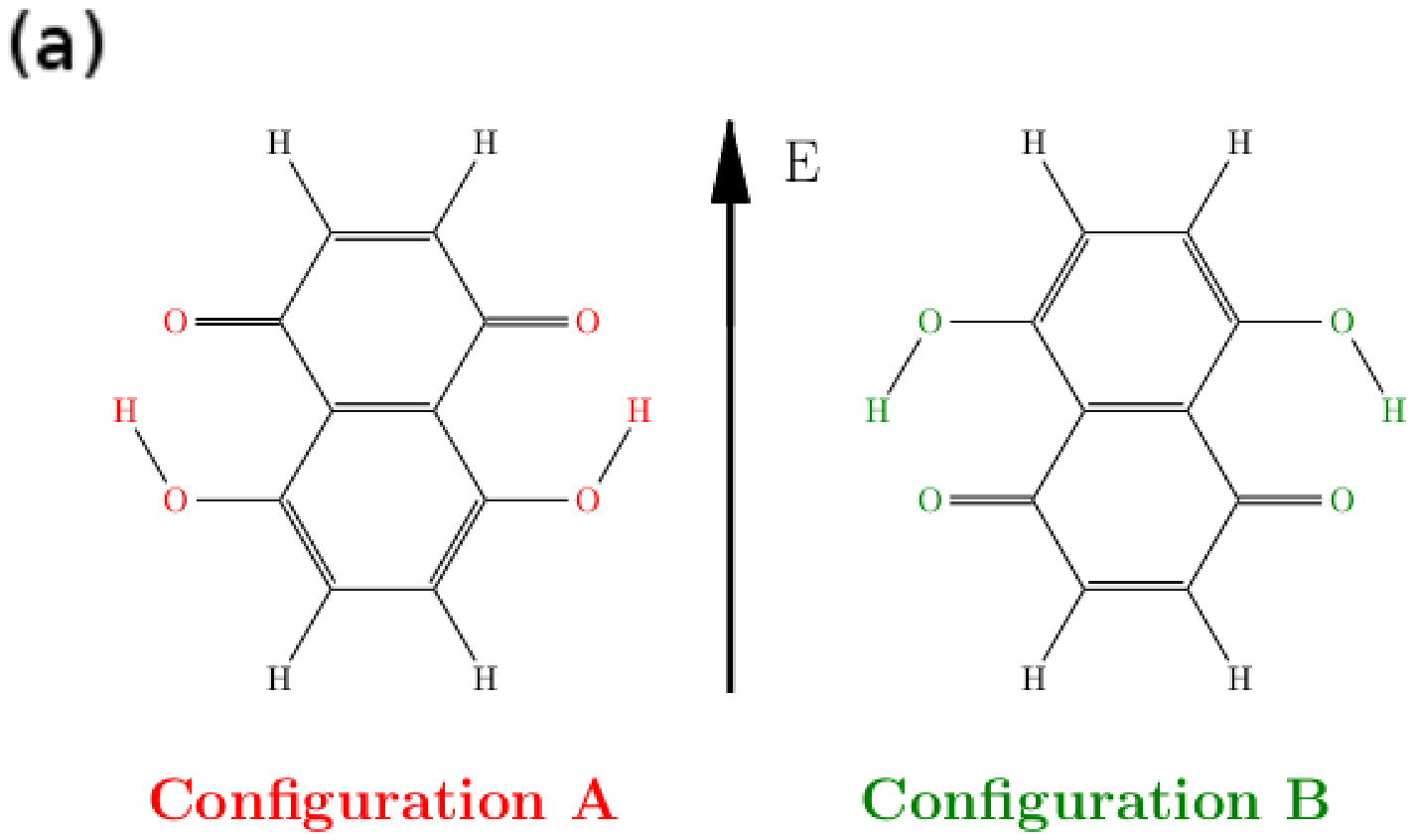} \hspace{4mm}
\includegraphics[scale=0.44,angle=0.0]{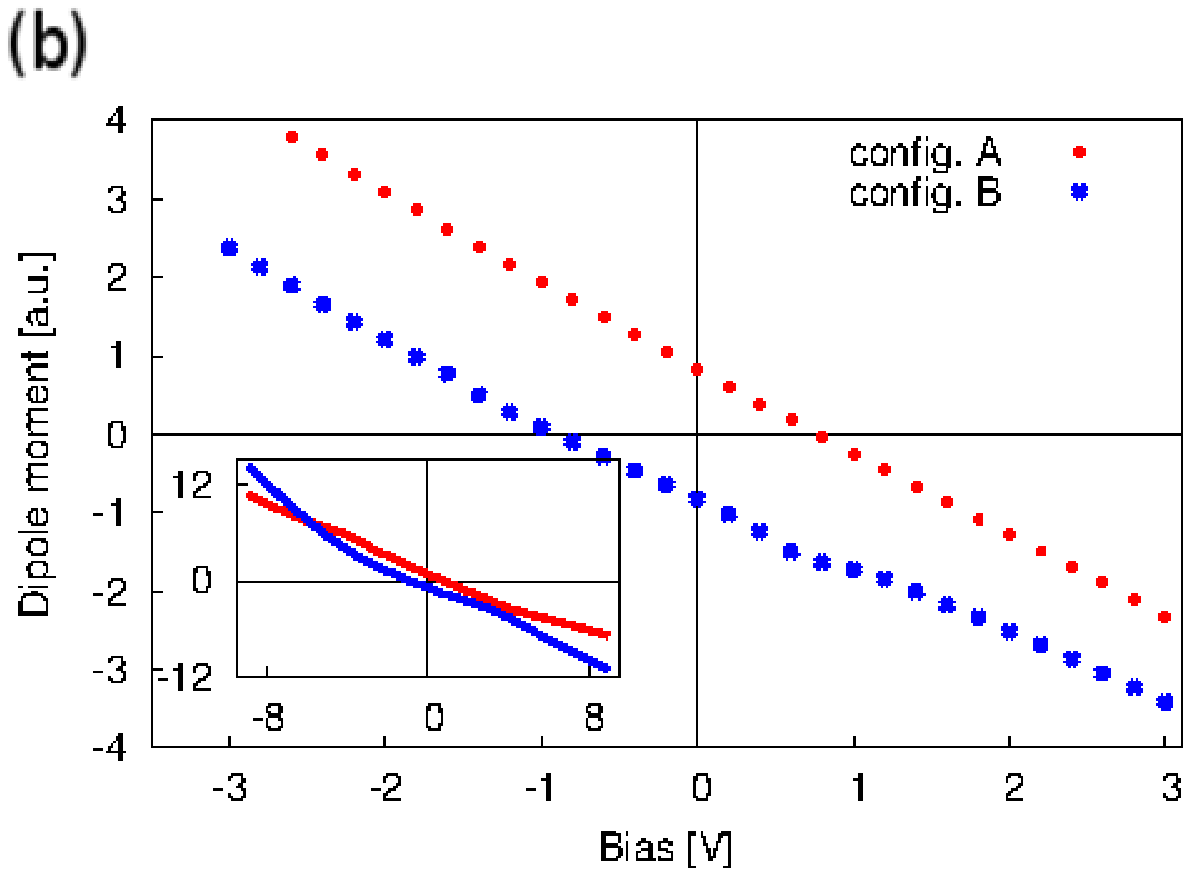} \hspace{2mm}
\includegraphics[scale=0.23,angle=0.0]{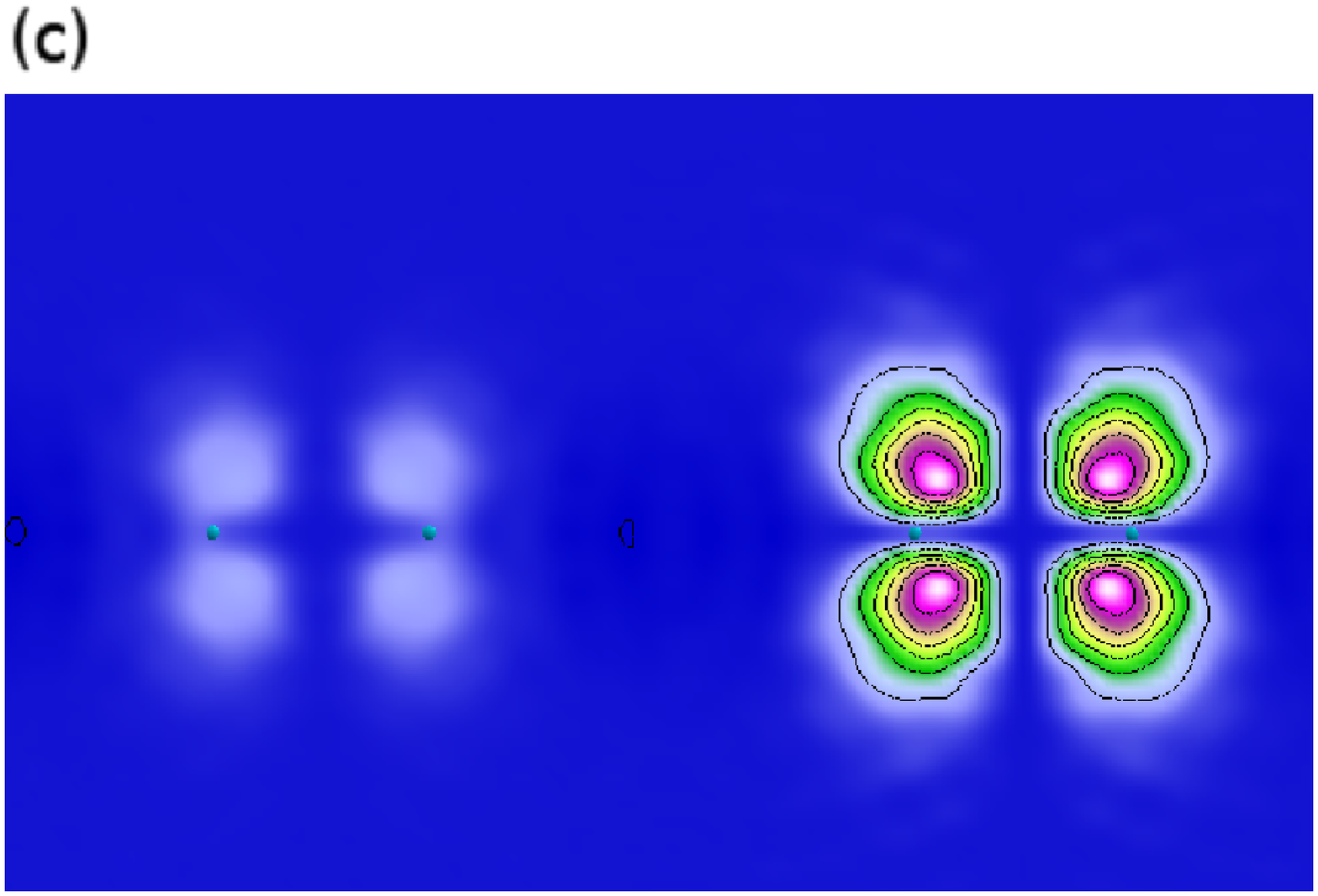} }
\caption{The proton transfer configurations:
(a) the geometries of the naphtazarin based 1-unit long molecules,
(b) the electric-field induced dipole moment in the 2-units long molecules, 
inset shows the larger range of the electric field,
(c) the top STM images of the 2-units long molecules, 
for the configuration A (left) and B (right);
the tip potential was 0.2 eV and the distance from the top hydrogens was 0.2 $\AA$.}
\label{mol}
\end{figure*}

In this work, new memory devices are theoretically proposed by means of
the density functional calculations, performed to describe the effects involved
in the reading and writing of the information.
The proceeding section, devoted to the results, is organized as follows:
First, a presentation of the molecules, being in  the origin of our idea,
and their properties is done. Further, the energetic
characterization of the proton transfer in some model systems is shown. 
In the third part, the analysis for various cases of the Z-GNR edge-saturation 
and the corresponding local magnetizations induced at the edges is performed.
At the end of the results section, the technical details of the calculations are given. 

\section{Results}

\subsection{Molecules with the proton transfer}

\begin{figure}[ht]
\includegraphics[scale=0.19,angle=0.0]{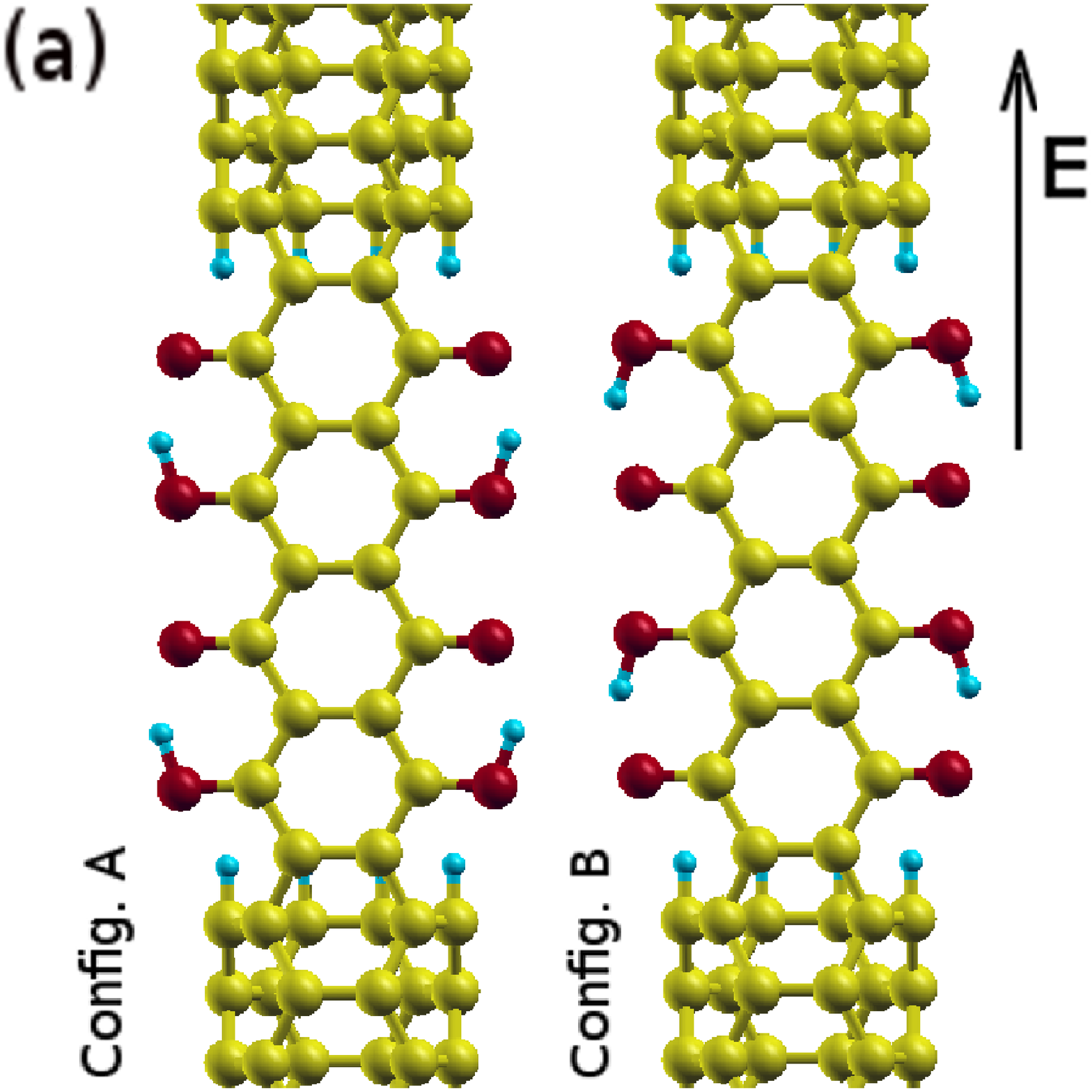} 
\vspace{8mm}
\includegraphics[scale=0.50,angle=0.0]{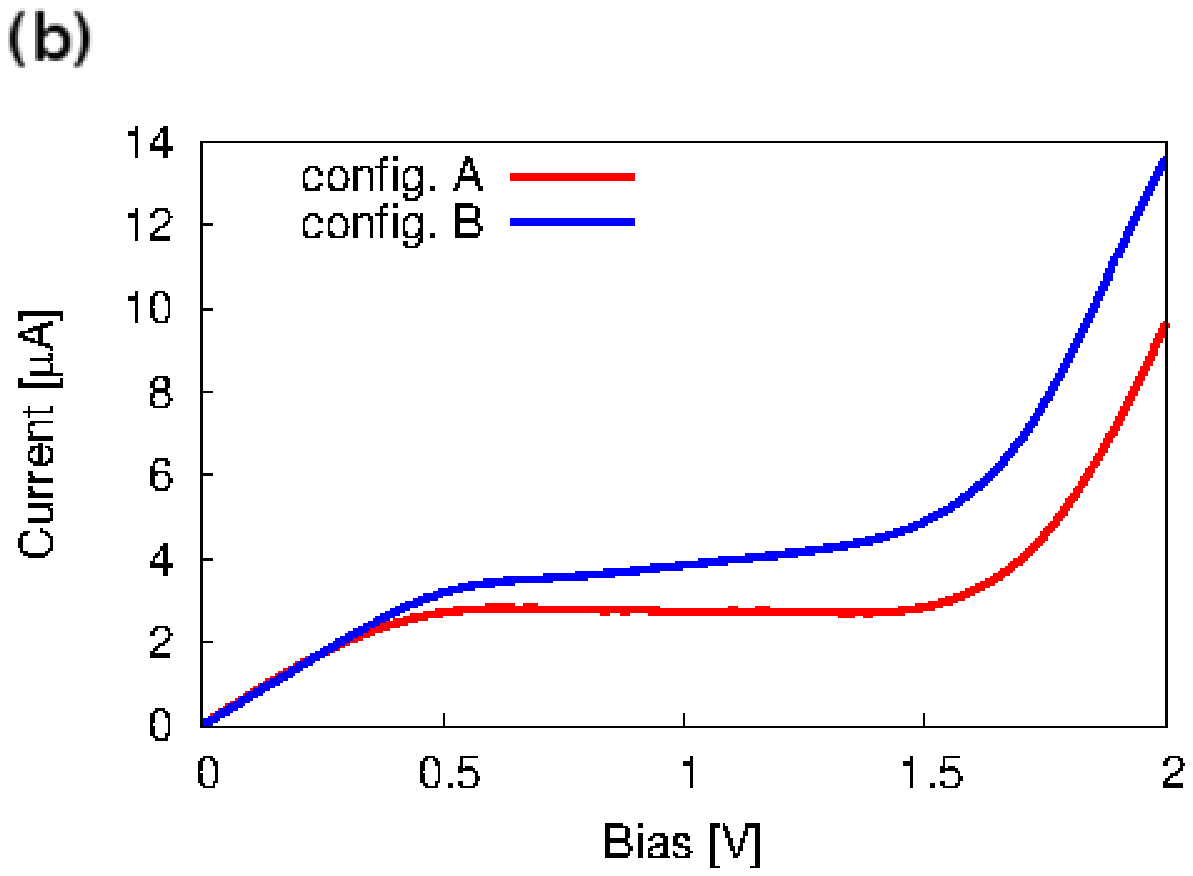} 
\caption{The 2-units long molecules attached to the carbon nanotubes:
(a) the geometry,
(b) the current-voltage characteristics.}
\label{IV}
\end{figure}

The proton transfer in the molecular systems is schematically presented in Fig. 1(a)
for two geometric configurations (A and B), which differ from each other by 
the location of the 'mobile' hydrogen atoms. 
These configurations are called '1-unit long', because 
one proton-hop O-H...O pair exists symmetrically on the two
sides of the axis parallel to the permanent dipole moment of the molecule.  
The two configurations (A and B) possess the dipole moments opposite to each other.
Applied electric field additionally polarizes the molecules and induces the corresponding
component of the dipole moment. Figure 1(b) shows the induced dipole moment for 
two configurations (A and B) of the 2-units long molecules 
versus the the applied voltage. 
These molecules could be used as memory units, if it was possible to detect 
the dipole moment in the direct way. 

The organic systems and honeycomb carbon structures, such as graphene and the carbon
nanotubes (CNT), 
are characterized by a network of the alternated single- and double-bonds. 
This alternation also characterizes the configurations A and B (in the opposite phase) 
\-- oxygen is connected by the double bond and the OH group by the single bond.
Therefore, the proton transfer induces changes in the alternation of the bond order.
For the symmetric systems, it means swapping of the configurations along the axis
parallel to the dipole moment. From the top view, the configurations
A and B are terminated with the horizontal double- or single-bond branch, respectively.
These details can be seen in the STM images; such maps are presented in Fig. 1(c) 
for the 2-units long molecules. The STM imaging is, however, not a good method
to detect the type of the PT configuration, because the applied electrostatic potential 
might induce the PT process; the measured logic state is switched
\-- depending on the sign of the STM-tip potential    
and the height of the energetic barrier for the proton transfer.

The proton-transfer molecules, 2-units long, attached to the CNT(6,0) are presented
in Fig. 2(a) for two configurations, A and B.
These cases are characterized by different  I-V curves, reported in Figure 2(b).
The current gear of the two configurations is, however, not pronounced well enough
to be used in the logic devices operating at low-bias conditions \-- 
the high voltage will change the measured logic state,
in the same way as the high potential of the STM-tip.
 
\subsection{Energy barriers and writing the logic state}

\begin{figure}[ht]
\includegraphics[scale=0.38,angle=0.0]{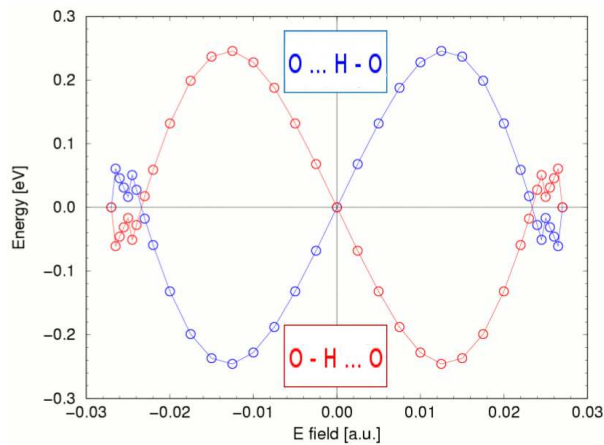}
\caption{Field-dependent relative energy differences between two configurations
of the 2-units long molecules (the orientations with respect to the electric 
field are marked with colors). Zero energy has been set to the mean energy of 
both structures at a given value of the field. Courtesy of A. L. Sobolewski.}
\label{bar}
\end{figure}

The total energies of the conformers, A and B, of the symmetric molecules 
are the same in the absence of the external electric field. In contrast, 
in the applied electric field,  the two configurations
of the 2-units long molecules are energetically distiguished;
Figure 3 presents the energy values relative to the total energy at the zero field. 
The configuration whose dipole moment is parallel to the field vector has lower energy.
The energy difference between the two configurations grows with the electric field,
up to some critical value of the E-field (with maximum of the energy separation between A and B).
Additionally, the E-field polarizes the molecules, which is the source of nonlinear effects
at higher values. The energies of the two configurations become closer again and swap 
by successive transfer of protons at the second critical value of the E-field \--
this is the next zero-energy crossing point. 

Detection of the logic states of the molecules - switched by the proton-transfer - 
is possible indirectly via the effect of the local magnetization, 
induced in the system with the frustrated network of the single- and double-bonds.
Such frustration can be generated in the zigzag graphene nanoribbon, which is saturated 
differently at the opposite edges, 
or there is an alternation of the chemical termination along the same edge.
The edges might end for example with oxygen or the OH group. 
A change of the edge saturation might be realized via the proton transfer 
from the edge ended with OH group to the STM-tip appex decorated with a small 
aromatic molecule with one H replaced by O; 
and vice versa.
Recently, it is possible to move a small molecule from 
one site at the surface to some other adsorption site, using the STM tip.
This can be done by changing the potential during the vertical and 
horizontal movements\cite{penseta-1,penseta-2}.  In the same way, it is also possible
to add or remove hydrogen to/from the graphene edge. 

\begin{figure}[ht]
\includegraphics[scale=0.25,angle=0.0]{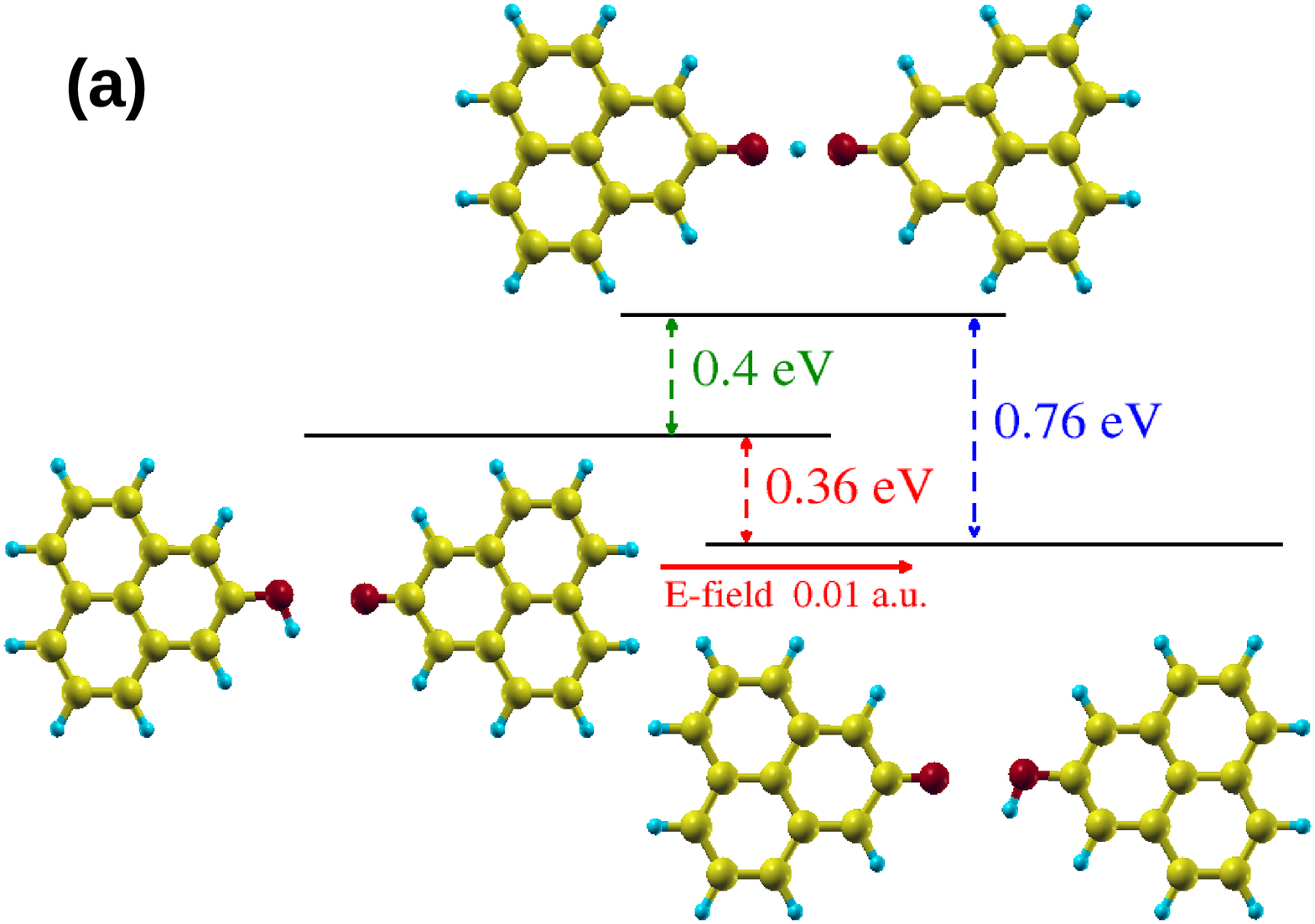}
\includegraphics[scale=0.25,angle=0.0]{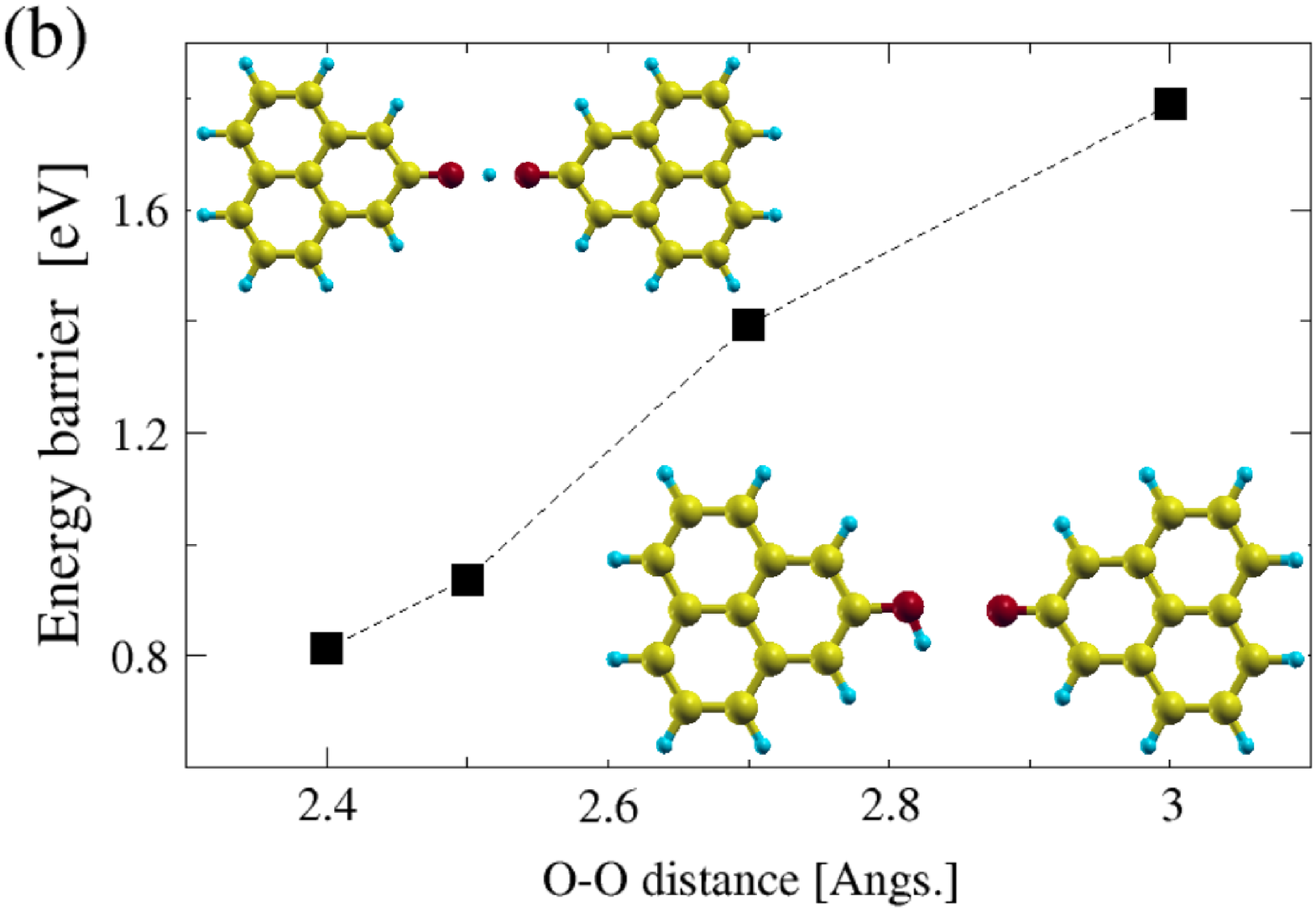}
\caption{Barriers for the proton transfer in the model structures
\-- a part of the zigzag GNR and a molecule attached to the STM appex:
(a) at the fixed distance of 2.4 $\AA$ between the oxygens,
(b) for the transition state at the zero field and as a function of the distance.}
\label{bar}
\end{figure}

\begin{figure*}[ht]
\centerline{
\includegraphics[scale=0.40,angle=0.0]{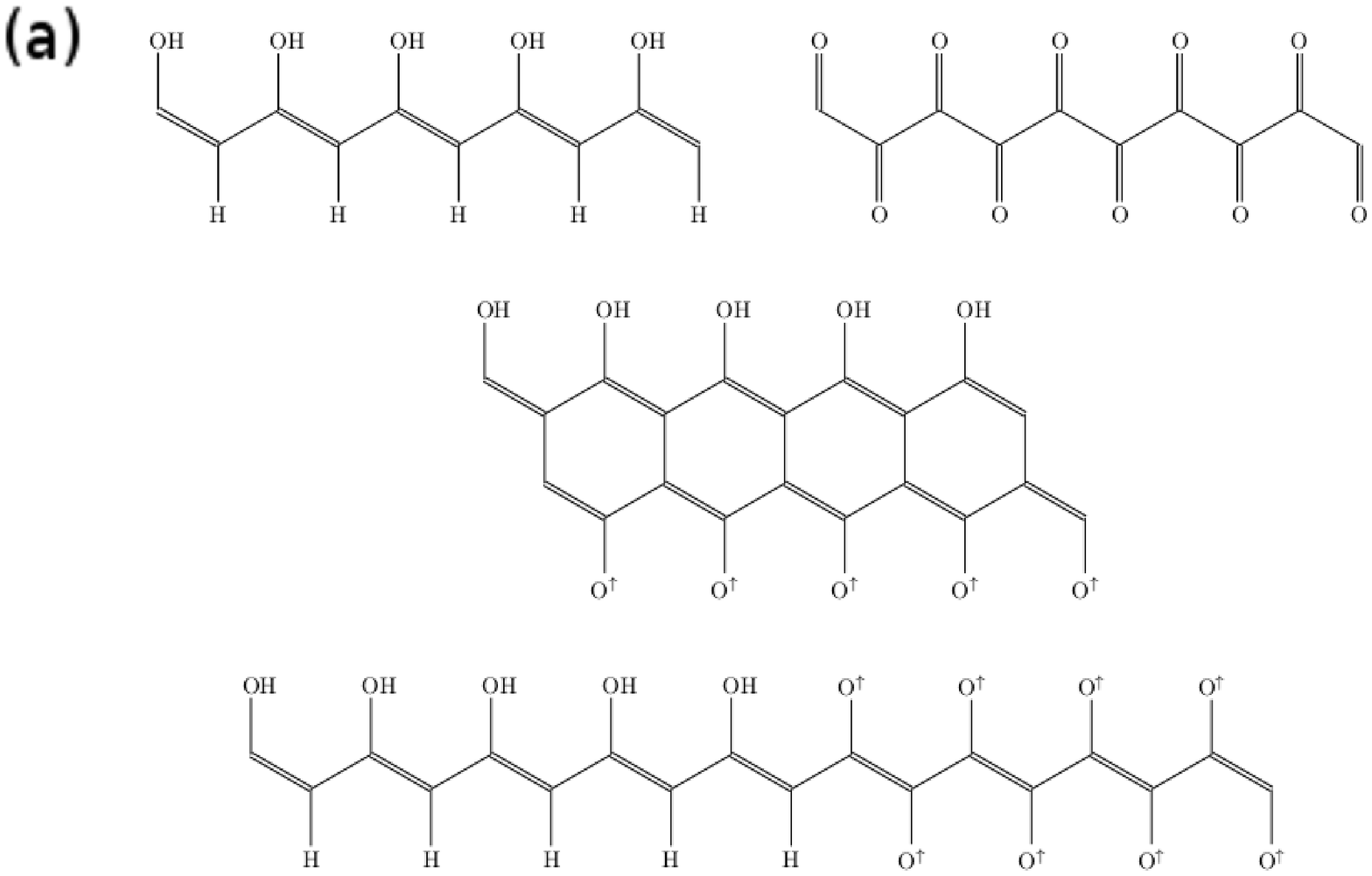}
\includegraphics[scale=0.25,angle=0.0]{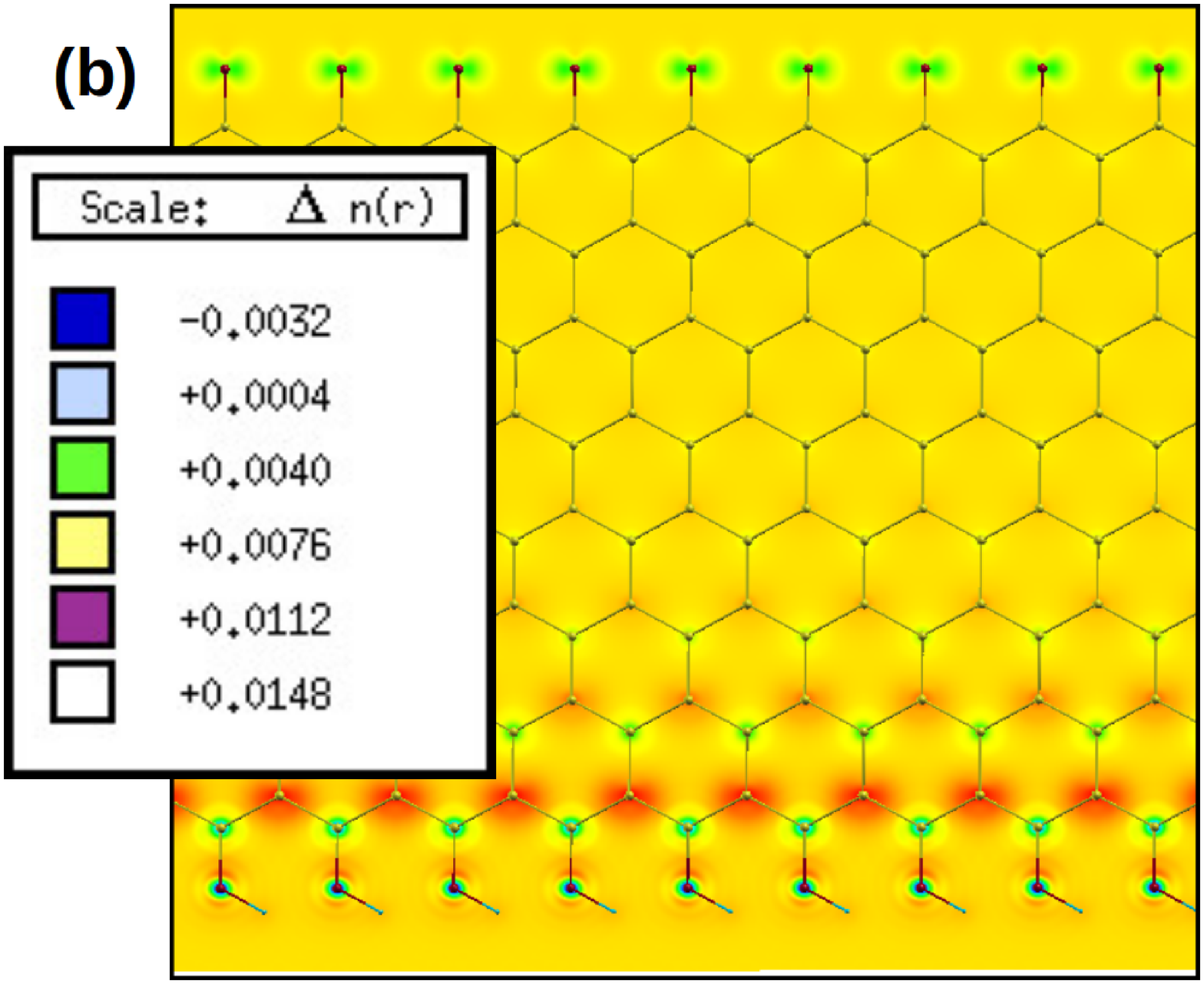}}
\centerline{
\includegraphics[scale=0.25,angle=0.0]{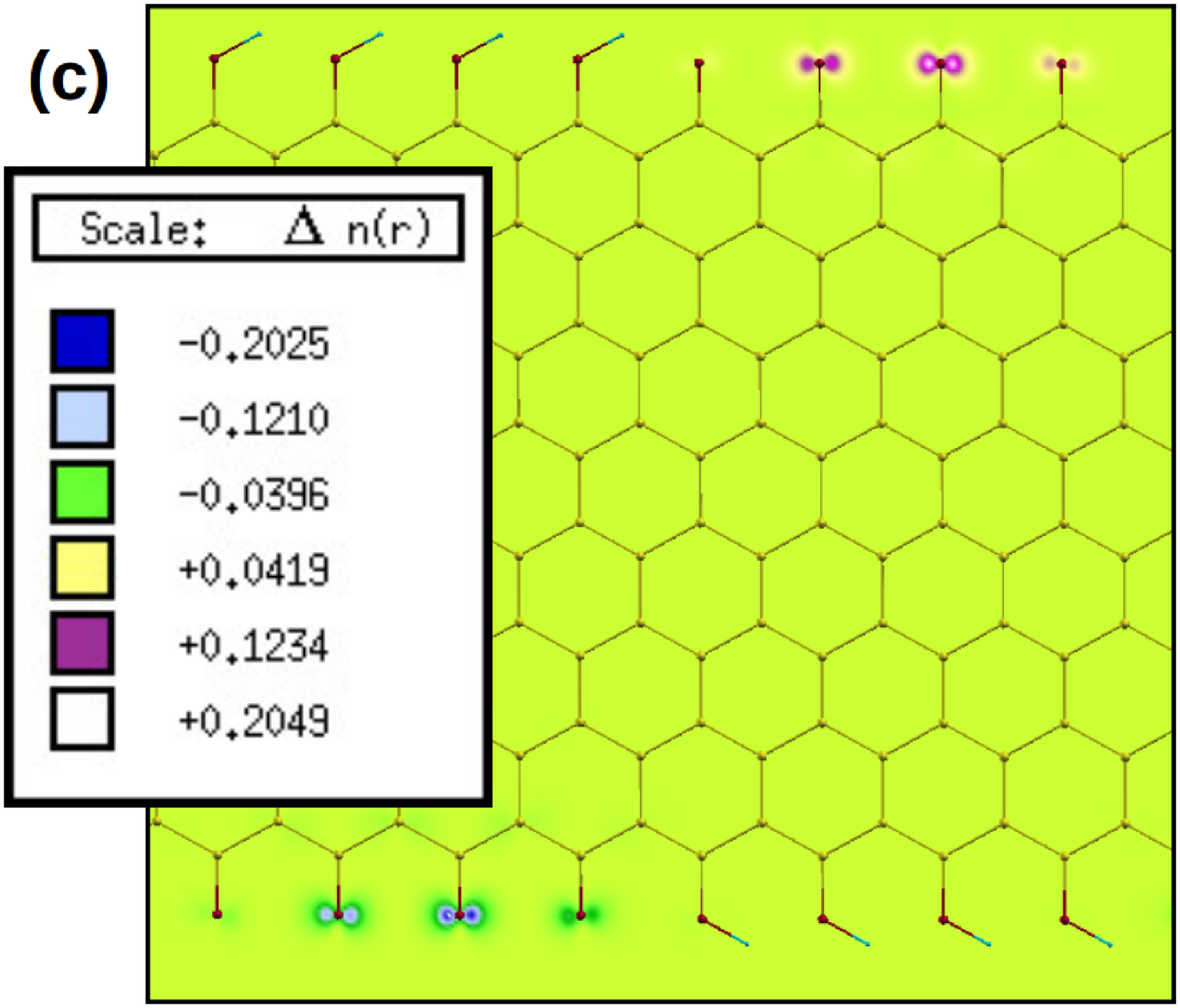}
\includegraphics[scale=0.25,angle=0.0]{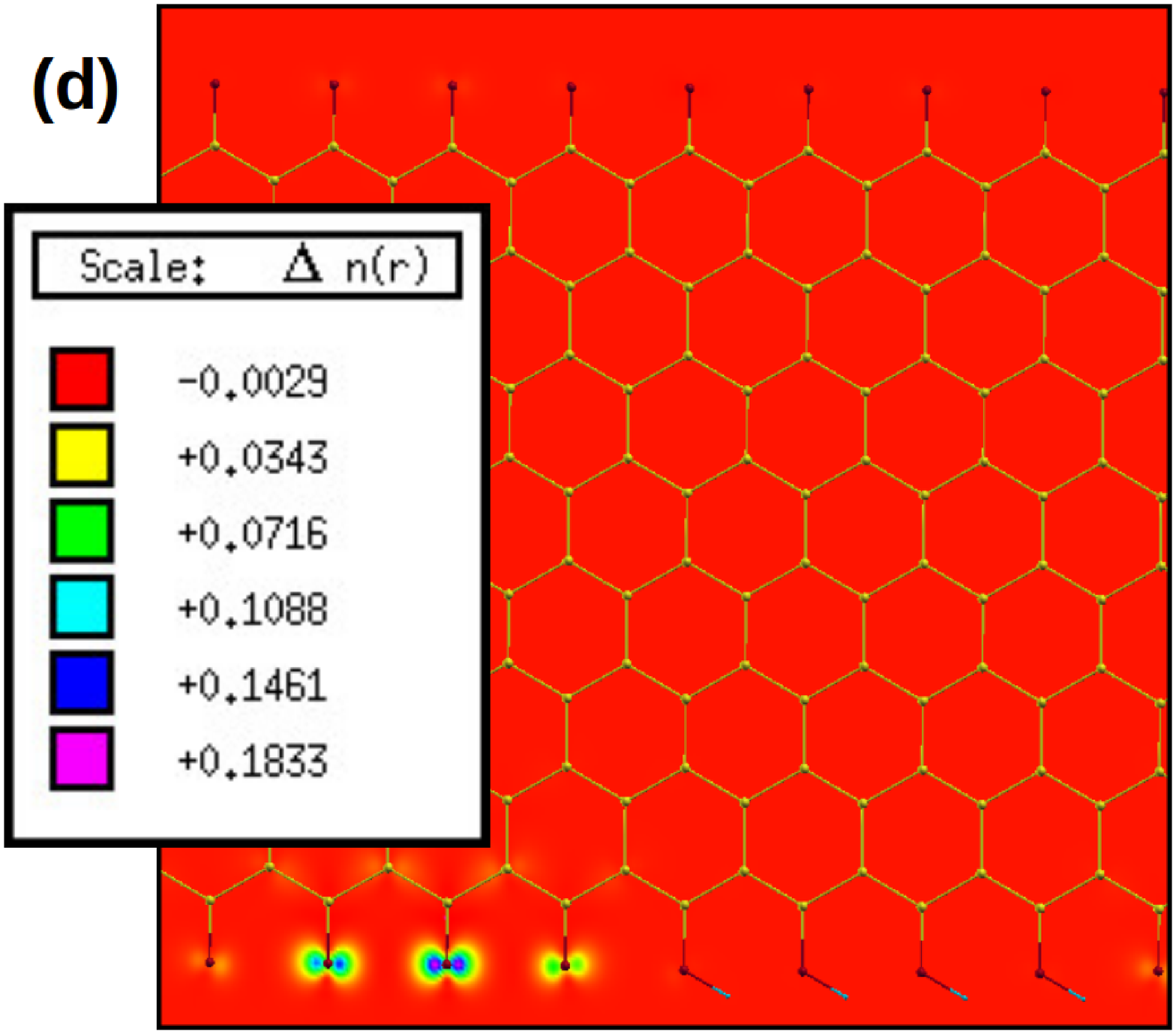}
\includegraphics[scale=0.25,angle=0.0]{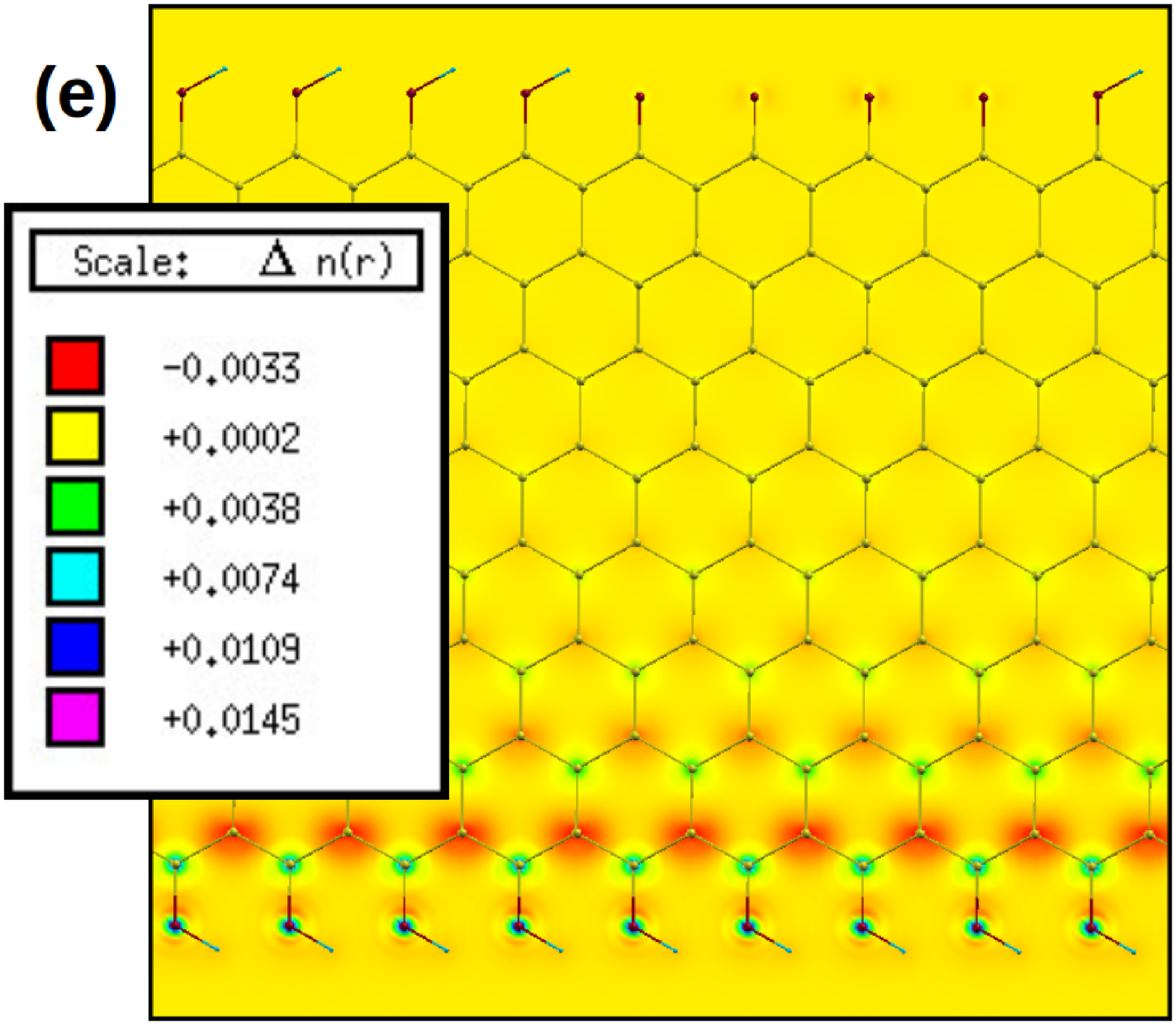}}
\centerline{
\includegraphics[scale=0.28,angle=0.0]{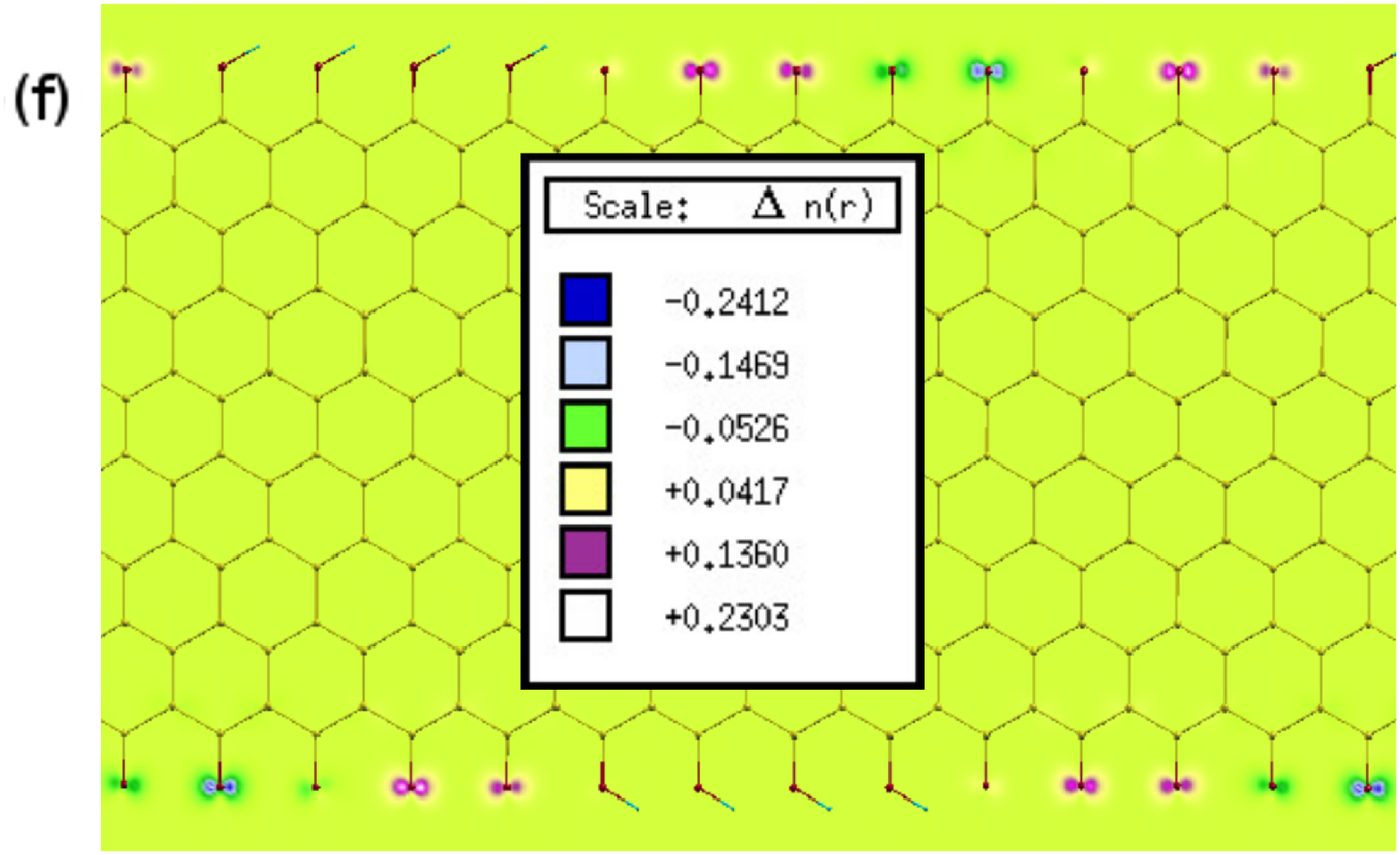}
\includegraphics[scale=0.25,angle=0.0]{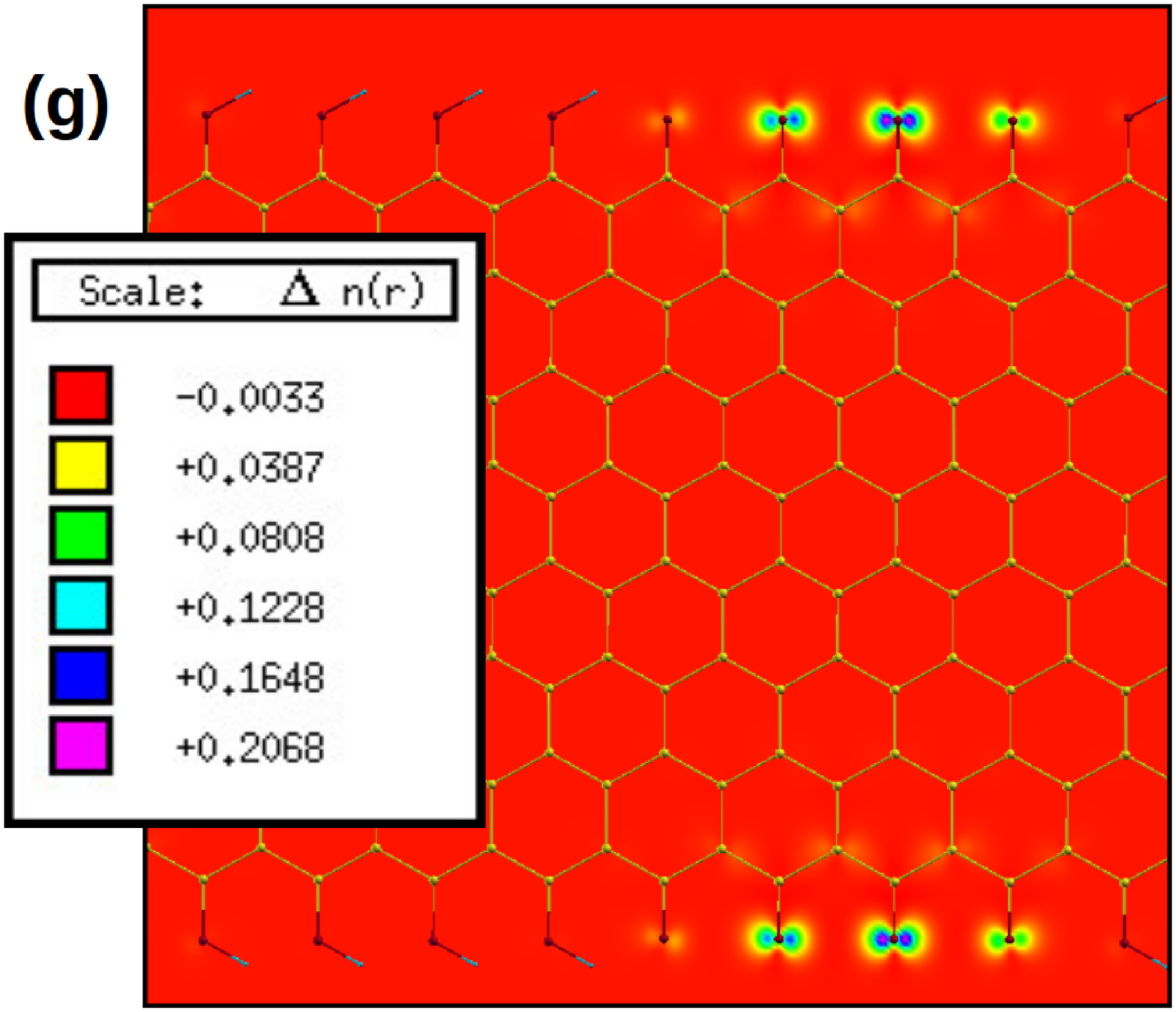}}
\caption{Edge magnetization in the Z-GNR;
(a) the scheme of the single- and double-bond network for the diffrent edge saturations,
and the spin-polarization maps:
(b) ferrimagnetism in the (OH...O) case,
(c) AF-coupling at the opposite O-edges in the (OH...O)+(O...OH) case,
(d) FM state at the (O+OH)-edge in the (O...O)+(O...OH) case,
(e) ferrimagnetism at the OH-edge in the (OH...OH)+(O...OH) case,
(f) AF-alternation along the (OH+O+O)-edge in the (O...OH)+(O...O)+(OH...O) case,
(g) FM state at the both O-edges in the (O...O)+(OH...OH) case. 
$\Delta$n denotes the spin up and spin down density difference}.
\label{mag}
\end{figure*}

The energy barriers for such PT process between the Z-GNR edge and the STM-tip molecule
are schematically printed in Figure 4(a). The left-hand side molecule represents a part of the
zigzag GNR with the edge saturated with the OH group. 
The right-hand side represents an aromatic molecule
attached to the STM tip, and it binds the oxygen atom which is able to capture proton.   
In the electric field oriented towards the STM tip, the total energy of this model system 
is lowered  when hydrogen is bound to the molecule at the STM-tip (right side), 
and not to the Z-GNR edge (left side).
The energetic barrier to be conquerred for the hop forward is defined as the energy 
difference between the transition state configuration 
\-- which is the one with the proton in the middle 
of the way between the Z-GNR and the STM tip (the top configuration) 
\-- and the initial state, with the proton bound to the Z-GNR edge (left side).
The barrier for the hop backwards is the energy difference between the transition state 
and the final state (right side). It is higher than the barrier for the forward process, 
which makes the system to prefer the one-way reaction.

Tuning the hop-forward and hop-backwards barriers 
is possible via changes of the distance between the Z-GNR edge and the STM-tip appex
\-- reported in Fig. 4(b) for the vanishing electric field 
\-- and also by varying the STM-tip potential.
For the non-volatility, it is better to chose the larger distance between the
Z-GNR edge and the STM-tip appex and the higher electric field, rather than the closer
distance and smaller field. This is because the distance increases the barriers between various 
configurations, and the electric field increases the difference between the initial and final
configuration (thus, also between the hop-forward and hop-backward barriers).  

Writing in the memory device could operate as follows: we start with the STM-tip
possessing the OH group at the attached molecule, and we point it to the site at the Z-GNR edge which
is saturated with O. Switching the potential at the tip to the positive value will move H to the Z-GNR.
The molecule at the tip is now saturated with pure O. We keep the positive potential at the tip, such
that none of the hydrogens from the Z-GNR can jump to the tip. We move the tip to the Z-GNR site from
which we want to take H, and we switch the tip potential to the negative value \-- such that,
the H atom jumps to the tip. Now, the STM-tip state is the same as it was on the start, and it can be
used again to add the H atom to the Z-GNR site which is saturated with O. We do it by moving the tip
and keeping the negative potential on the way, and we change the potential to the positive value 
when the tip points towards the desired O atom at the Z-GNR.

Since the STM device operates on the samples in vacuum, the 
well chosen distances and potentials should give enough guaranty for the correct coding.
In every moment, the tip can be used for only one type of the operation: either adding or removing
the H atom to/from the Z-GNR edge. Therefore, one needs to move the tip along the edge forward and
backward \-- when it is necessary \-- keeping the potential set to the value, which prevents the H jump
(i.e. such that the hopping barrier is higher), while changing the tip's position. 
The current state of the tip is always dependent on the previous operation. The time for the
proton transfer from the tip is similar to that of moving H to the tip. 
This is because the aromatic molecule at the STM-tip appex is similar to the graphene edge,
from the saturated atom point of view. 

The impractical point is that the STM machines are big devices. On the other hand,
such device  could be simplified by removing the part 
which is responsible for making the images of the surface 
\-- since we only need to precisely move the tip and change the potential.
The future of these devices might go towards the downsizing. 

\subsection{Magnetic reading from the memory units}

Kondo magnetism in graphene with defects is well established.\cite{Kondo} 
The nature of the defected 'bulk' nanoribbon states in comparison
to the metallic edge-states was also studied.\cite{nl-Cantele} 
The effect being in focus of this work, namely the Z-GNR edge magnetization, 
has been discovered by Fujita et al.\cite{mag-1} 
Recently, it has been proposed to use this phenomenon for the spin filtering
of the current, flowing along the Z-GNR junction with
differently saturated edges.\cite{spin-filter,filter2}

Figure 5(a) schematically shows the mechanism of the Z-GNR edge magnetization.
When the both edges are saturated via the single bonds, then the double bonds in  
the network alternate the single bonds along the zigzags parallel to the edges. 
In the reversed situation, when both edges are saturated via the double bonds,
then the double bonds in the Z-GNR network alternate the single bonds 
along the zigzags perpendicular to the edges. Different kind of the saturation at 
the opposite edges, or a change of the chemical termination along the same edge, 
introduces a frustration in the Z-GNR. Therefore, the electrons,
which should form the double-bonds at the edge, are released from the chemical bonds  
due to the above effect of the frustration, and they are involved in the local magnetization. 

The spin maps of various edge-saturation cases are presented in Figures 5(b)-5(g).
In the homogeneously saturated Z-GNR with one edge ended with O and
the opposite one with the OH group \-- in Fig. 5(b) \-- the ferrimagnetic state forms. It characterizes
with the magnetic moments localized at oxygens and a kind of 'spin-wave' at the OH-edge.
The total magnetic moment per the elementary cell 
(cell with 19 atoms in the zigzag across the GNR width) is
0.28 $\mu_B$ and the absolute magnetic moment is 0.47 $\mu_B$.
If the above case is modified such that the O-saturated edge is alternated with
the OH saturation every four units \-- presented in Fig. 5(e) for the 8-units cell \-- 
then the magnetic moment at the O-edge segments is quenched.
Although, the total and absolute magnetic moments
per unit are similar to the earlier case \-- since now these values are 2.26 $\mu_B$ and 3.95
$\mu_B$, respectively, per 8-units long cell.
When the first case saturation with O and OH at the opposite edges is modified by
a 'flip' of the edge-satuartion in every fourth unit \-- as it is in Fig. 5(c) \-- 
then only the O-saturated ends possess the local moments.
These local moments at the opposite edges couple to each other antiferromagneticaly (AF)
\-- with a very little energetic gain, of 1 meV, with respect to the
ferromagnetic (FM) coupling. The absolute moment per 8-units cell is 2.27 $\mu_B$ in the both
AF and FM cases, and the total moment in the FM case is 1.76 $\mu_B$.

Seems that in all cases, the single- and double-bond alternation along the edge is never frustrated.
If one edge is totaly O-saturated and the opposite edge saturation is alternated with O and OH at every
fourth unit, then only one edge (that with the O-ended parts between the OH-ended segments) 
shows the local magnetic moment and only at oxygens \-- in Figure 5(d). 
The total and absolute magnetizations per 8-units long cell in this case
are 0.77 $\mu_B$ and 1.01 $\mu_B$, respectively. The L\"owdin population analysis shows
a little bit smaller charge at the magnetized oxygens with respect to the nonmagnetic ones, 
about 0.3 in the spin minority.
Similar situation occurs for the alternation (at every fourth unit) 
of the OH-OH both-edge terminated units 
and the O-O both-edge saturation, in Fig. 5(g). 
Oxygens on both edges possess the local magnetism, with the total and absolute values
of 1.8 $\mu_B$ and 2.27 $\mu_B$, respectively, per 8-units cell.
When the longer segments of the oxygen saturation (8-units) reside between the OH-saturated
segments (4-units) along the same edge (in the 12-units long elementary cell) 
and these OH-parts are not symmetrically placed on the opposite edges \-- in Fig. 5(f) \--
the AF-coupled alternation forms along the O-segments.
The total and absolute magnetizations, in this larger cell,
achieve the values of 1.65 $\mu_B$ and 5.33 $\mu_B$, respectively.

Rich diversity of the magnetic orderings, existing at the edges of the Z-GNR under the 
inhomogeneous saturation, enables to construct various boolean operations. 
Reading the logic states could be performed using the giant-magnetoresistance 
device.\cite{mem-gmr} 
For instance, the simplest way of reading the linear coding, would be 
to move the GMR device along one edge containing the segments with O and OH, while keeping the other 
edge completely saturated with O \-- this is the case in Fig. 5(d).
In contrast, if the edge opposite to the side read by the GMR is totaly saturated with the OH groups
only, there is almost no signal on the segment-alternated edge; see Fig. 5(e). 
The manipulation on both sides of the Z-GNR is more complicated and needs two GMR devices
simultaneously reading the local magnetization at both edges.  

Due to the fact, that the system is closed in vacuum conditions, 
we do not have any gas reservoir of the hydrogen atoms -
the number of bits 0 and bits 1 seems to be limited in the device.
In order to create a source of additional H and the storage space for not needed H, 
we can use another Z-GNR. Thus, one Z-GNR would be an element of memory
and the other should be saturated with O in half of its length and O-H in the second half.
When adding or taking H from this Z-GNR reservoir, a marker of the last position for 
the saturation-change point is necessary. The room temperature effect for writing 
is very small (of about 0.025 eV at 300 K)
and does not affect the energy barriers for the H transfer, which are 10 times larger.

The proposed device can operate as the magnetic random access memory (MRAM), 
due to the high density, non-volatility and fast writing and reading processes - if one 
uses the GMR or magnetic tunnel junction (MTJ) or 
spin-transfer torque (STT) head for reading. 
Moreover, our way of writing is not limited by the large write currents in the way 
of Oersted-field flipping in magnetic-type writing. These currents increase 
the power consumption well beyond that of
the static and dynamic RAM.\cite{rev,sramdram} 

For reading purposes, it is convenient to introduce separators between 
the bits 0 and 1. Such separators would be of the same length as one bit and consist of 
the Z-GNR edge saturated with O and O-H. 
Thus, the reading device needs to move the head to the right bit's position. 
In such case, all bits 1 would be magnetised in the same way, independently of 
the length of a sequence of bits 1 and the neighboring bit; the same for bits 0. 
Then, reading the bits 0 would not be any problem, because the measured current 
is the same as in the case of zero magnetic field. To read the bit 1, one needs  
ro use a sensitive head, 
which is able to detect the current difference at the magnetic field of 1 Bohr magneton 
at room temperature. In order to deal with a decrease of the tunnel magnetoresistance 
current at elevated temperatures, one could use for instance the Heusler compounds 
as reading heads.\cite{rev}

\subsection{Theoretical details}

The molecular calculations for the energy barriers in the external electric field
were performed with the quantum chemistry package TURBOMOLE.\cite{turbo}
The wavefunctions are represented, 
in the code, in the localized basis set;
our choice was the correlation-consistent valence double-zeta Gaussian basis
set with the polarization functions for all atoms (cc-pVDZ).\cite{basis}
The computational method was the density functional theory\cite{dft} (DFT) 
in the generalized gradient approximation (GGA) scheme. 
The functional has been chosen for B3LYP,\cite{b3lyp}
which is a combination of the DFT and 20$\%$ of the exact exchange \--
necessary for qualitatively accurate description of the hydrogen bonds.

The calculations of the graphene-edge magnetization were perfomed 
with the {\sc Quantum ESPRESSO} (QE) package.\cite{qe} 
These codes use the plane-wave basis set 
and the pseudopotentials to describe the core electrons.
The exchange-correlation functional was chosen for the Perdew-Burke-Erzenhoff (PBE) 
parametrization.\cite{pbe} 
The ultrasoft pseudopotentials have been used, with the energy cutoffs of 30 Ry and 240 Ry 
for the plane-waves and the density, respectively.
For the edge-magnetization calculations in the Z-GNR, 
the Monkhorst-Pack uniform k-mesh in the Brillouin zone (BZ) has been set to
$80\times 1\times 1$ for the 1-unit cell, $40\times 1\times 1$ for the 8-units cell, 
and $20\times 1\times 1$ for the 12-units cell.  
The Fermi-surface energy broadening parameter of 0.005 Ry
has been chosen for a better convergence.
The vacuum separation between the periodic slabs was 20 $\AA$.

The dipole moment and the current-voltage curves have been obtained with the 
Wannier-functions postprocessing tools.\cite{w90} 
Firstly, the DFT band-structure calculations were performed. Further,
the maximally-localized Wannier functions have been obtained.\cite{mlwf,RMP}
In the end, the dipole moment has been calculated according to the modern
theory of polarization.\cite{dip-1,dip-2} 
The ballistic transport calculations were performed according to 
the Landauer-B\"uttiker approach using the transport tool\cite{trans} 
within the wannier90 suite of codes. 

The spin-polarization maps and the STM image have been prepared using 
the XCrySDen package.\cite{tone} \\

\section{Conclusions}

In this paper, new random access and flash memory devices are proposed by means of 
the first-principles predictions. 
The basic concept relies on the proton transfer from the OH-saturated Z-GNR edge 
into the STM tip decorated with a molecule, 
which binds O, and vice versa.
Such processes are very fast, and their energetics can be tuned by the distance
between the tip and the Z-GNR edge, as well as the electrostatic potential. 
Thus, enabling picosecond and non-volatile writing of the information
to the logic cells.

The proton transfer process reverts the local dipole moment and changes the pattern of 
the single- and double-bond network in the organic and carbon honeycomb systems.
This mismatch between variously patterned networks of the chemical bonds, in turn, 
causes the local magnetic moment in the Z-GNR with the edges saturated
differently on the opposite sides or along the same side. 
The induced magnetic moment can be read quickly by the GMR device. \\

{\bf Acknowledgements} \\
Prof. A. L. Sobolewski is kindly acknowledged for 
the discussions and Figure 3.
This work has been supported by 
The National Science Center of Poland
(the Project No. 2013/11/B/ST3/04041).
Calculations have been performed in the Interdisciplinary Centre of
Mathematical and Computer Modeling (ICM) of the University of Warsaw
within the grant G59-16. \\

\end{document}